\documentclass[11pt]{article}

\usepackage[utf8]{inputenc}
\usepackage{cite}  % in text chain sequential citations in bibliography with a dash
\usepackage{graphicx}
\usepackage{amsmath}
\usepackage{subcaption}
\usepackage[margin=1.25in]{geometry}
\usepackage[usenames,dvipsnames]{color}
\usepackage{url}
\usepackage[colorlinks = true,
            linkcolor = blue,
            urlcolor  = blue,
            citecolor = blue,
            anchorcolor = blue]{hyperref}

\usepackage{authblk}

\usepackage{lineno}

\newenvironment{Abstract}{\begin{quotation} \begin{center}
                       ABSTRACT
     \end{center}\bigskip  }{\end{quotation}}

\newcommand\snowmass{\begin{center}\rule[-0.2in]{\hsize}{0.01in}\\\rule{\hsize}{0.01in}\\
\vskip 0.1in Submitted to the  Proceedings of the US Community Study\\ 
on the Future of Particle Physics (Snowmass 2021)\\ 
\rule{\hsize}{0.01in}\\\rule[+0.2in]{\hsize}{0.01in} \end{center}}

\def\beq{\begin{equation}}
\def\eeq#1{\label{#1}\end{equation}}
\def\eeqn{\end{equation}}

\newenvironment{Eqnarray}%
   {\arraycolsep 0.14em\begin{eqnarray}}{\end{eqnarray}}
\def\beqa{\begin{Eqnarray}}
\def\eeqa#1{\label{#1}\end{Eqnarray}}
\def\eeqan{\end{Eqnarray}}

\def\lsim{\mathrel{\raise.3ex\hbox{$<$\kern-.75em\lower1ex\hbox{$\sim$}}}}
\def\gsim{\mathrel{\raise.3ex\hbox{$>$\kern-.75em\lower1ex\hbox{$\sim$}}}}

\def\del{\partial}
\def\Dslash{\not{\hbox{\kern-4pt $D$}}}
\def\dslash{\not{\hbox{\kern-2pt $\del$}}}
\def\pslash{\not{\hbox{\kern-2pt $p$}}}
\def\ETmiss{\not{\hbox{\kern-4pt $E$}}_T}

\def\Dlr{\mathrel{\raise1.5ex\hbox{$\leftrightarrow$\kern-1em\lower1.5ex\hbox{$D$}}}}

\def\MSB{{\bar{M \kern -2pt S}}}
\def\msb{{\bar{\scriptsize M \kern -1pt S}}}

\def\drb{{\bar{\scriptsize D \kern -1pt R}}}

\title{Graph Neural Networks in Particle Physics: Implementations, Innovations, and Challenges}
\date{\vspace{-5ex}}

\author[1]{Savannah Thais\thanks{Contact Editor, sthais@princeton.edu}}
\author[2]{Paolo Calafiura}
\author[3]{Grigorios Chachamis}
\author[1]{Gage DeZoort}
\author[4]{Javier Duarte}
\author[5]{Sanmay Ganguly}
\author[6]{Michael Kagan}
\author[2]{Daniel Murnane}
\author[7]{Mark S. Neubauer}
\author[6]{Kazuhiro Terao}

\affil[1]{Princeton University}
\affil[2]{Lawrence Berkeley National Lab}
\affil[3]{Laborat{\' o}rio de Instrumenta\c{c}{\~ a}o e F{\' \i}sica Experimental de Part{\' \i}culas (LIP)}
\affil[4]{University of California San Diego}
\affil[5]{ICEPP, University of Tokyo}
\affil[6]{Stanford Linear Accelerator Laboratory}
\affil[7]{University of Illinois at Urbana-Champaign}

\begin{document}

\maketitle

\snowmass{}
\begin{Abstract}
Many physical systems can be best understood as sets of discrete data with associated relationships. Where previously these sets of data have been formulated as series or image data to match the available machine learning architectures, with the advent of graph neural networks (GNNs), these systems can be learned natively as graphs. This allows a wide variety of high- and low-level physical features to be attached to measurements and, by the same token, a wide variety of HEP tasks to be accomplished by the same GNN architectures. GNNs have found powerful use-cases in reconstruction, tagging, generation and end-to-end analysis. With the wide-spread adoption of GNNs in industry, the HEP community is well-placed to benefit from rapid improvements in GNN latency and memory usage. However, industry use-cases are not perfectly aligned with HEP and much work needs to be done to best match unique GNN capabilities to unique HEP obstacles. We present here a range of these capabilities, predictions of which are currently being well-adopted in HEP communities, and which are still immature. We hope to capture the landscape of graph techniques in machine learning as well as point out the most significant gaps that are inhibiting potentially large leaps in research.
\end{Abstract}

\clearpage

%-------------------------------------------------------------
\section{Introduction}
\label{sec:introduction}
%-------------------------------------------------------------
Machine learning (ML) has had a profound impact on physics research; particularly in particle physics, ML has been successfully applied to a broad range of critical tasks including data collection, physics object reconstruction and identification, Standard Model (SM) measurements and new physics searches, experiment design and operation, and more \cite{ml_physics_community_whitepaper,modern_ml_particle_physics,ml_physics_living_review}. Initially, ML applications in particle physics focused on traditional classification and regression methods (boosted decisions trees, support vector machines, shallow neural networks (NNs), etc) using physics-motivated high-level features. However, more recent work has employed a variety of more complex deep learning architectures including deep NNs, convolutional neural networks (CNNs), and recurrent neural networks (RNNs). These methods allow the use of low-level information, often energy deposits in detectors, rather than derived variables and have inspired an assortment of different data representations such as images and sequences. Additionally, the adoption and integration of cutting-edge ML methods has enabled closer collaboration between the particle physics and ML communities and created opportunities for physics researchers to directly contribute to the development of state-of-the-art ML architectures.

Over the past several years, Geometric Deep Learning (GDL) has emerged as a highly impactful sub-field of ML focused on learning from non-Euclidean data structures including sets, groups, graphs, and manifolds \cite{gdl_overview}. These methods have been successfully applied to a broad range of scientific and societal domains like knowledge dependency representation \cite{knowledge_graphs}, physical system modeling \cite{battaglia2016interaction}, chemical and drug discovery \cite{drug_discovery_gnn_summary}, community connection mapping \cite{community_gnns}, and much more. In particular, graph neural networks (GNNs) are an impactful class of GDL algorithms that operate on graphs. Excellent reviews and summaries of GNNs detailing the spectrum of current implementations are available in \cite{gdl_overview,battaglia2018relational,gnn_overview1,gnn_overview2,gnn_overview3} and we provide a introductory overview of GNNs in the following section.

\begin{figure}
    \centering
    \includegraphics[width=6in]{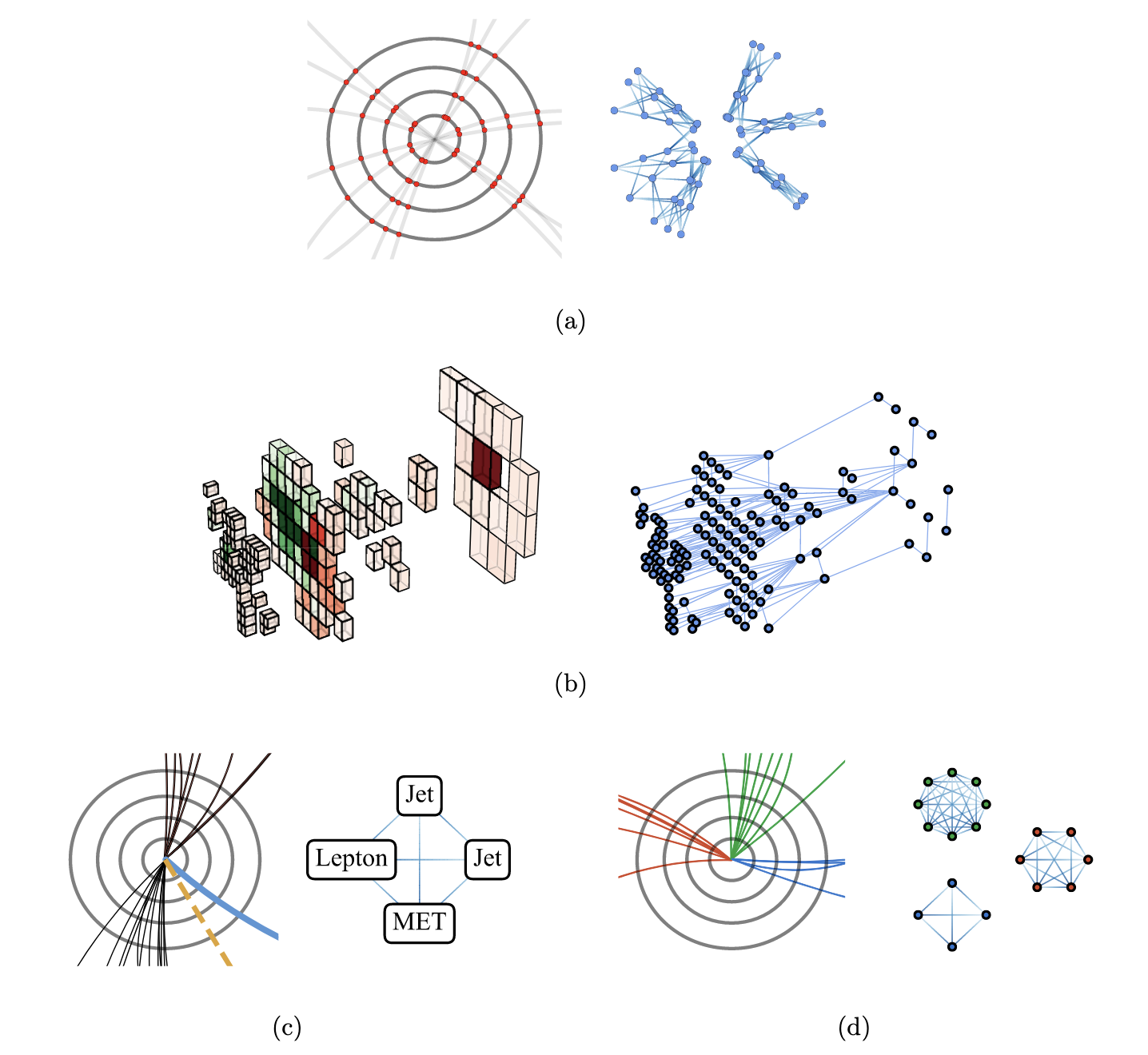}
    \caption{Examples of graph representations of particle physics data: (a) clustering tracking detector hits into tracks, (b) segmenting calorimeter cells, (c) classifying events with multiple types of physics objects, (d) jet classification based on the particles associated to the jet. Image taken from \cite{gnns_in_particle_physics}}
    \label{fig:graph_examples}
\end{figure}

Graphs are a data representation that describes objects (represented as graph nodes) and their pairwise relationships (represented as graph edges). Graph structures are able to effectively capture complex relationships and dependencies between objects which is essential for accurately representing physical data. For particle physics data, graph-based representations provide several advantages over alternative data representations: unlike vector- or grid-like structures, graphs allow for variable size data (i.e. one does not have to lose information or zero-pad structures); additionally, graphs are better suited for dealing with sparse and heterogeneous detector data that can be difficult to project into image-based representations and do not require the application of an artificial ordering scheme as required by sequence-based representations. Graphs are able to represent a broad range of particle physics data including energy deposits in a detector, individual physics objects like tracks or missing energy, individual particles or groups of particles, or even heterogeneous information; an example of different graph representations of particle physics data is shown in Figure \ref{fig:graph_examples}. 

Given the advantages of graph-based representations, it is unsurprising that GNNs have been successfully applied to many problems in particle physics. Many of these applications, particularly GNNs applied to data reconstruction tasks, are summarized in \cite{gnns_in_particle_physics,gnn_tracking_recon} and a range of current GNN applications are described in Section \ref{sec:current_uses}. However, despite the success of these methods, there are still substantial challenges that must be overcome in order to maximize the potential of these techniques and integrate them into particle physics experiments. In addition to providing an overview of current implementations, this paper focuses on describing these outstanding challenges and considerations specific to physical applications of GNNs in Section \ref{sec:challenges} and makes suggestions for directions researchers should focus on to address these challenges in Section \ref{sec:future}.

\subsection{Graph Neural Networks}
Following the success of convolutional neural networks for grid-structured data (i.e. pixels in an image), graph neural networks aim to extend many of the same powerful techniques to irregular, graph-like data structures. Initial work was done in parallel on generalising recurrence \cite{1555942, 4700287, 5596796, li2015gated, pmlr-v80-dai18a} and convolution \cite{bruna2014spectral, DBLP:journals/corr/HenaffBL15, DBLP:journals/corr/KipfW16} operations to graphs, so-called RecGNNs and ConvGNNs, respectively. Historically, RecGNNs were motivated by time-series data and therefore handled graphs that were dynamic across time but had little sophistication in how relational data was communicated. Conversely ConvGNNs were motivated by spectral graph theory \cite{chung1997spectral, shuman2013emerging}, using the graph Laplacian to capture higher-order features, in analogy with the filtering of features performed by CNNs. In recent years however, research has converged on some common, high-performing conventions. Convolutions have been shown to perform very well in the ``spatial" domain \cite{micheli2009neural, atwood2016diffusion, niepert2016learning, gilmer2017neural} - that is, not necessarily in physical space, but as opposed to the spectral domain. A spatial convolution is generally defined by two steps: an aggregation step and an update step. There is some arbitrariness to this distinction, but for most GNN tasks that involve learning some hidden representation of graph nodes, it is useful. 

\begin{figure}
    \centering
    \includegraphics[width=0.7\linewidth]{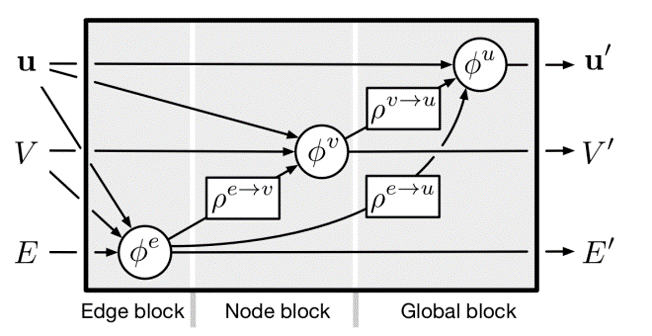}
    \caption{A GNN convolution as defined by \cite{DBLP:journals/corr/abs-1806-01261}. $\phi^e$ defines a message function, $\rho^{e\rightarrow v}$ is an aggregation around nodes, and $\phi^v$ is a node update. Graph-level features $\textbf{u}$ can optionally be produced.}
    \label{fig:GN_block}
\end{figure}

We begin by defining a graph $G = (\textbf{u}, V, E)$ as a collection of node features $\{v_i\} = V$, edge features $\{e_{ij}\} = E$, and graph features $\{u_g\} = \textbf{u}$. For the $(l+1)^{th}$ convolution iteration, a node $i$'s hidden representation $v^{l+1}_i$ can be computed by
\begin{align*}
    & Aggregate: & v^{l'}_i &= \rho(e^{l+1}_{ij}), && \textnormal{where} && e^{l+1}_{ij} = \phi^e(v^{l}_i, v^{l}_j, e^{l}_{ij}) \quad \textnormal{and} \quad j\in \mathcal{N}_i\\
    & Update: & v^{l+1} &= \phi^v(v^{l'}_i, v^{l}_i, u^l)
\end{align*}
In words: Features are aggregated around a node $i$'s neighborhood $\mathcal{N}_i$ by first computing each $e_{ij}$, called the ``message" on the edge connecting node $i$ and node $j$. The particular message function $\phi^e$ is dependent on the choice of architecture, and may be isotropic (a node treats all of its neighbors equivalently) or anisotropic (a node has some mechanism of attention, or edges have their own feature space). The message function may also include an MLP. Messages are aggregated around a node, where $\rho$ stands for any permutation-invariant aggregation. The new node features can be combined with previous node features, or some higher features $u$ belonging to the graph (or even node-level, non-local features), and passed through a node-wise MLP. This node update is represented by $\phi^v$. We can represent the whole process by the diagram in fig \ref{fig:GN_block}, which also allows for graph features $\textbf{u}$.

\begin{figure}[htb!]
    \centering
    \begin{subfigure}[a]{0.8\linewidth}
        \centering
        \includegraphics[width=\textwidth]{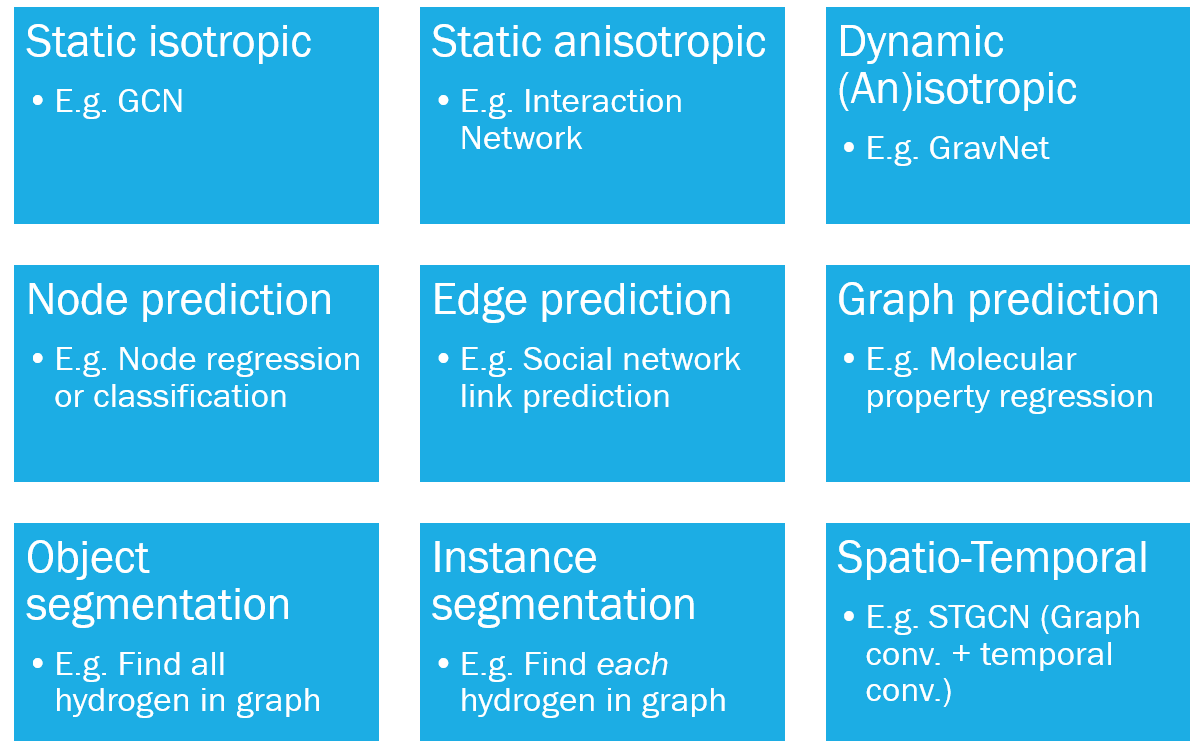}
        \caption{The landscape of GNNs}
        \label{fig:GNN_landscape}
    \end{subfigure}
    \vspace{3em}
    \begin{subfigure}[b]{0.8\linewidth}
        \centering
        \includegraphics[width=\textwidth]{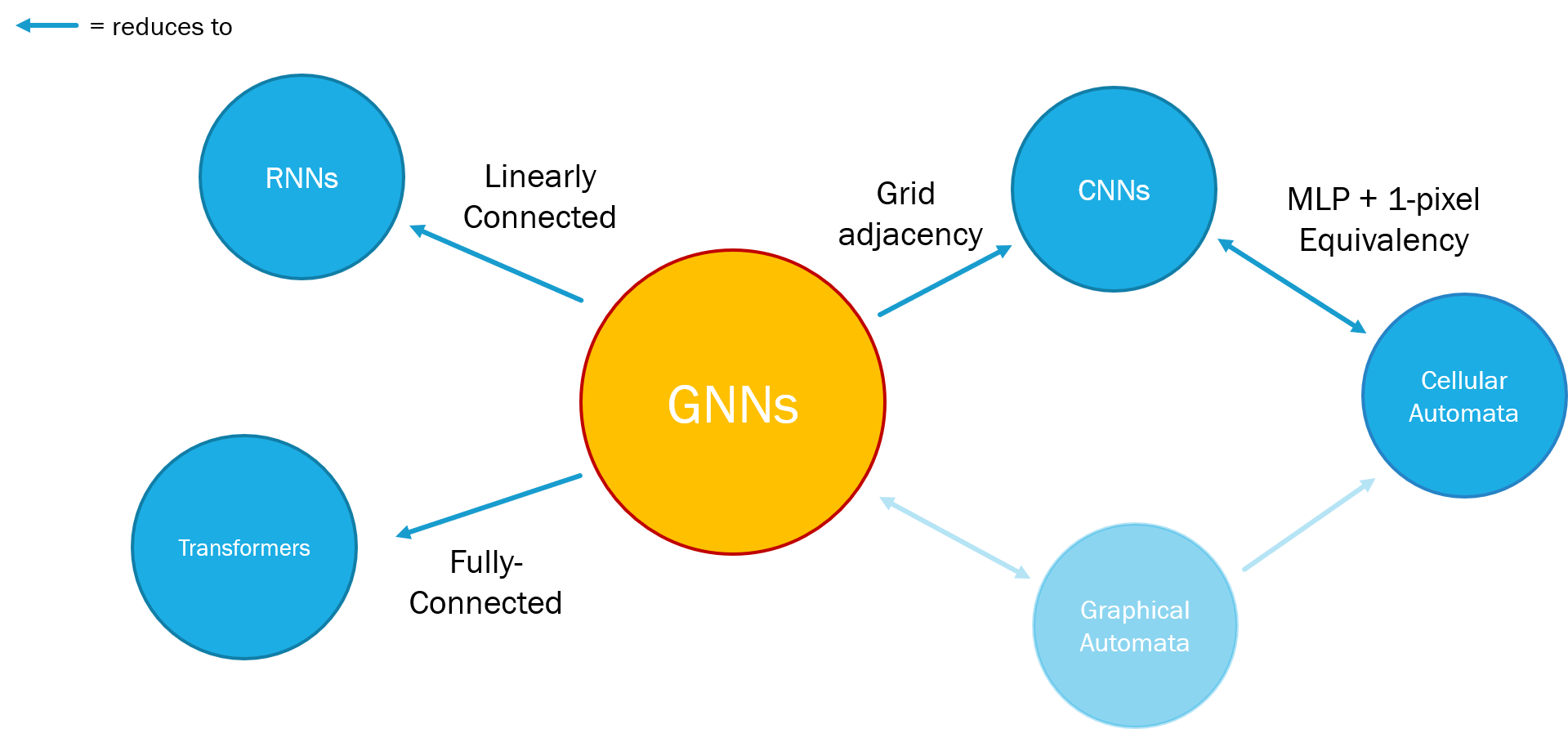}
        \caption{Reduction properties of a GNN. }
        \label{fig:GNN_reduction}
    \end{subfigure}
    \label{fig:GNN_landscape_and_reduction}
    \caption{GNN taxonomy}
\end{figure}

Just as GNN convolutions have settled on a general language and set of best-practices, so has the notion of recurrence. Because GNNs now typically have multiple convolutions (or ``message passing steps"), there are two dimensions for feature update: one across message passing steps (i.e. aggregation and update MLPs share weights), and one across spatio-temporal steps (i.e. each MLP shares weights with a previous time-step). The former is common, and generally good practice for reducing the size of a GNN and improving training stability. The latter is an area of active research in so-called Spatio-Temporal GNNs \cite{song2020spatial, DBLP:journals/corr/abs-2006-10637}, where edge connections and even the existence of nodes may change over time. In the case of edge connections changing between message passing steps, we will call these Dynamic GNN architectures. The landscape of typical GNNs is given in fig \ref{fig:GNN_landscape}.

%-------------------------------------------------------------
\section{Current Uses of GNNs in HEP}
\label{sec:current_uses}

%-------------------------------------------------------------
\subsection{Reconstruction and Identification}
\label{sec:reconstruction_and_identification}

To date, the majority of graph-based learning algorithms in HEP focus on reconstruction (clustering) and identification (classification) tasks. Despite having fundamentally different objectives, their corresponding GNN-based workflows have structural similarities. Most begin with an unordered set of data points, for example tracker hit spatial positions or particle-level kinematic features. Graph construction routines embed these points as graph nodes, extending edges between them to represent relational information. This process may occur explicitly before the GNN, or otherwise be repeated dynamically as part of the learning algorithm. The choice of edges has important downstream effects; message-passing GNNs learn by aggregating information across each node’s local neighborhood, which is defined by its incident edges. GNNs produce node-level, edge-level, or graph-level predictions; these predictions may be leveraged by a subsequent post-processing algorithm. In this way, graph-based learning algorithms can be characterized by their implementations of three key steps: 1) graph construction, 2) GNN inference, 3) post-processing. 

\textbf{Reconstruction} applications typically involve an underlying clustering task, which may occur in real space or in a learned latent space. Some of the first GNN-based reconstruction algorithms followed the \textit{edge classification} paradigm, in which GNNs are trained to predict the strength of each node-node relationship (an edge weight) encoded by an edge \cite{farrell_novel_2018}. This approach is prevalent in GNN-based tracking pipelines, where edges represent hypothesized particle trajectories; in this scheme, edge weights produced by a GNN may be used to reject edges outright to form a set of disjoint subgraph clusters, or leveraged by a downstream track clustering algorithm. Many variants of this pipeline have been proposed, including a variety of graph construction algorithms and post-processing track finding modules \cite{dezoort_charged_2021, biscarat_towards_2021, ju_performance_2021}. These edge-classifying GNN architectures are frequently based on the Interaction Network \cite{battaglia2016interaction}, iteratively applying both \textit{edge blocks}, MLPs designed to re-embed each set of edge features, and \textit{node blocks}, message passing modules that leverage the re-embedded edge features, to a latent representation of the input graph. 

Though the edge classification paradigm has been applied to calorimeter segmentation in HGCal-like data \cite{ju_graph_2020}, most GNN-based calorimeter reconstruction studies focus on node classification, such as predicting the fractional assignments of hits to different showers, and graph classification tasks like particle identification and energy regression. Many such applications leverage the novel GravNet and GarNet GNN layers  \cite{qasim_learning_2019}, proposed as lightweight alternatives to EdgeConv-style~\cite{edgeconv} dynamic graph construction and subsequent message passing. Notably, GravNet has been implemented on an FPGA via HLS4ML \cite{iiyama_application_2019}. GravNet was recently used to facilitate object condensation \cite{qasim_multi-particle_2021}, the process of clustering nodes belonging to a common object and extracting that object’s properties in one-shot, on HGCal-like calorimeter data \cite{kieseler_2020_object}.
EdgeConv-based networks have also been used for overlapping calorimeter shower disentanglement \cite{DiBello:2020bas}. Additionally, Dynamic Reduction Networks exist at the intersection of reconstruction and classification, learning an optimal graph pooling strategy (e.g. calorimeter hit clustering) to boost subsequent classification or regression performance~\cite{gray_dynamic_2020}. 

Event-level reconstruction tasks have also been explored by graph-based methods. 
For example, pileup rejection has been addressed as a node classification in which pileup scores are predicted for particles embedded as graph nodes~\cite{martinez_pileup_2019, abcnet}.  
A GNN-based PF algorithm was developed to operate on graphs with heterogeneous nodes corresponding to tracks and calorimeter clusters \cite{mlpf,Pata:2022wam}. 
Set2Graph functions, a class of learnable approximators mapping sets to graphs, were applied to vertex reconstruction by predicting edges between an input set of tracks \cite{Serviansky:2020qwa}. 

\textbf{Identification} tasks usually focus on graph classification or segmentation. Jet identification, the process of tagging a particle initiating a jet, has been addressed by a wide range of graph-based learning algorithms. Many focus on jets represented as particle clouds, unordered sets of particles embedded as graph nodes with kinematic features, subsequently applying set-based learning functions (e.g. Energy Flow Network)~\cite{deepsets, energyflow, dolan_2021_equivarEFN}, message passing with adjacency learning ~\cite{henrion2017neural}, or EdgeConv blocks~\cite{particlenet}. Others still represent jets as graphs explicitly, for example applying interaction networks~\cite{moreno_jedi-net_2020} or graph attentional pooling layers~\cite{abcnet} to produce classification scores. Jet substructure has been explicitly leveraged to identify jets, for example through secondary vertex finding~\cite{Shlomi:2020ufi} or graph-based models representations of two-point correlations between subjets~\cite{chakraborty_neural_2020}. Set-based classifiers have also been applied to jets represented as sets of tracks~\cite{atlas_collaboration_deep_2020}. 

GNNs have also been applied to signal identification, typically at the level of an entire event. 
For example, MPNNs have been applied to event graphs containing heterogeneous particle nodes (i.e. the node features contain explicit particle labels) with distance-weighted edges to classify stop pair production~\cite{abdughani_probing_2019}, CP odd/even Higgs decays to $b\bar{b}$ with associated semileptonic $t\bar{t}$ decays~\cite{ren_unveiling_2020}, and di-Higgs events with $b\bar{b}WW^*$ final states~\cite{abdughani_probing_2021}. Permutation invariant set-based architectures (with no explicit graph structure) have been developed to address the combinatorics of jet matching in fully-hadronic $t\bar{t}$ events with novel attention mechanisms~\cite{fenton_permutationless_2021, lee_zero-permutation_2020}. 
Graphs have also been used to represent decay chains in semi-leptonic $t\bar{t}$ decays, wherein particles are embedded as heterogeneous nodes and parent-child decay relationships are represented as edges~\cite{atkinson_improved_2021}. 

\subsection{Data Generation}
\label{sec:generative_modeling}
\textbf{Sanmay, Grigoris, Jean-roch}
%\TODO{add a little intro about ML based generators (VAEs, GANs, normalizing flows (https://arxiv.org/pdf/2105.09016.pdf?) ) then dive into how GNNs can be used, why data generation matters}

Generative models has a potential big impact in producing simulated data, to be used for high energy physics analysis \cite{Arjona_Mart_nez_2020}. A real detector simulation is extremely time consuming and CPU intensive task. In future, for high pileup run condition like HL-LHC, the complexity of a full simulation is going to scale-up extensively. Hence a generative model based fast simulation, which can reliably replicate the detector smearing and shower generation, is essential. 

The most general representation of a general collider event is a point-cloud or graph, where each node of a point-cloud refers to an energetic cell in the detector or a hit in the tracker. Hence a generative model for producing graph data is a natural choice for many high energy physics simulation tasks. 

%\TODO{move to new tasks}
There is another potential application of GNNs on even more theoretical grounds. Our current understanding of perturbation theory applied within the SM is mostly based on a diagrammatic approach. Although there have been studies that focus on the complexity and adjacency matrix representations of Feynman diagrams (mainly for integrable systems in Quantum Chromodynamics and $N = 4$ SYM theories) the use of GNNs has not yet been employed. As one introduces higher order corrections, which generally correspond to diagrams with more loops and more external legs, the number of diagrams to be computed increases very rapidly. However, we know that at each order in the perturbative expansion, there are unifying principles shared by the Feynman diagrams at that order that could be exploited and potentially lead to a much faster computation of the matrix elements needed to calculate partonic cross-sections. These are needed for almost every phenomenological study at collider experiments.

A message passing based point cloud generation has been suggested in article \cite{Kansal:2021cqp} where the authors propose a new architecture called MPGAN to solve the task. It is shown that the MPGAN model outperforms other generative models on every considered metric. Another auto-encoder based GNN has been proposed in the article \cite{hariri2021graph} to reconstruct LHC events. A great amount of progress in fast simulation has been achieved using generative model, as has been shown recently \cite{ATLAS:2021pzo}. Usage of message passing based networks will further enhance the performance in future. Another crucial task will be multi-segmented point cloud generation for collider events for future colliders.

\subsection{Algorithm Acceleration}
While large experimental workflows are typically run on CPUs, it can be advantageous to run ML inference on heterogeneous coprocessors, such as graphics processing units (GPUs), field-programmable gate arrays (FPGAs), or even specialized AI processors like Graphcore IPUs~\cite{graphcore,Maddrell-Mander:2020atc}, Intel Habana Goya~\cite{goya} and Gaudi~\cite{gaudi} cards, Google TPUs~\cite{tpu}, and Cerebras wafer-scale engines~\cite{cerebras}.
Using heterogeneous computing resources as a service for ML inference has been demonstrated for a variety of use cases in experimental workflows~\cite{Duarte:2019fta,Krupa:2020bwg,Rankin:2020usv}.
In particular, large GNN models like ParticleNet~\cite{particlenet} and machine-learned particle-flow (MLPF)~\cite{mlpf,explaining_mlpf,Pata:2022wam} have been accelerated using GPUs as a service.
Work has also been done to accelerate the inference of neural networks with FPGAs~\cite{FINN,FINNR,fpgadeep,fpgaover,Duarte:2018ite,Summers:2020xiy,bnnpaper,Coelho:2020zfu,Aarrestad:2021zos}, including GNNs~\cite{Iiyama:2020wap,Heintz:2020soy,Elabd:2021lgo,6861577,10.1145/3007787.3001155,10.1145/3373087.3375312,yan2020hygcn,9218751,geng2020awbgcn,kiningham2020grip}.
Acceleration of GNN training and inference on coprocessors is promising direction for future R\&D.

%-------------------------------------------------------------
\section{Challenges}
\label{sec:challenges}
\begin{figure}[htb!]
    \centering
    \begin{subfigure}[a]{0.8\linewidth}
        \centering
        \includegraphics[width=0.8\textwidth]{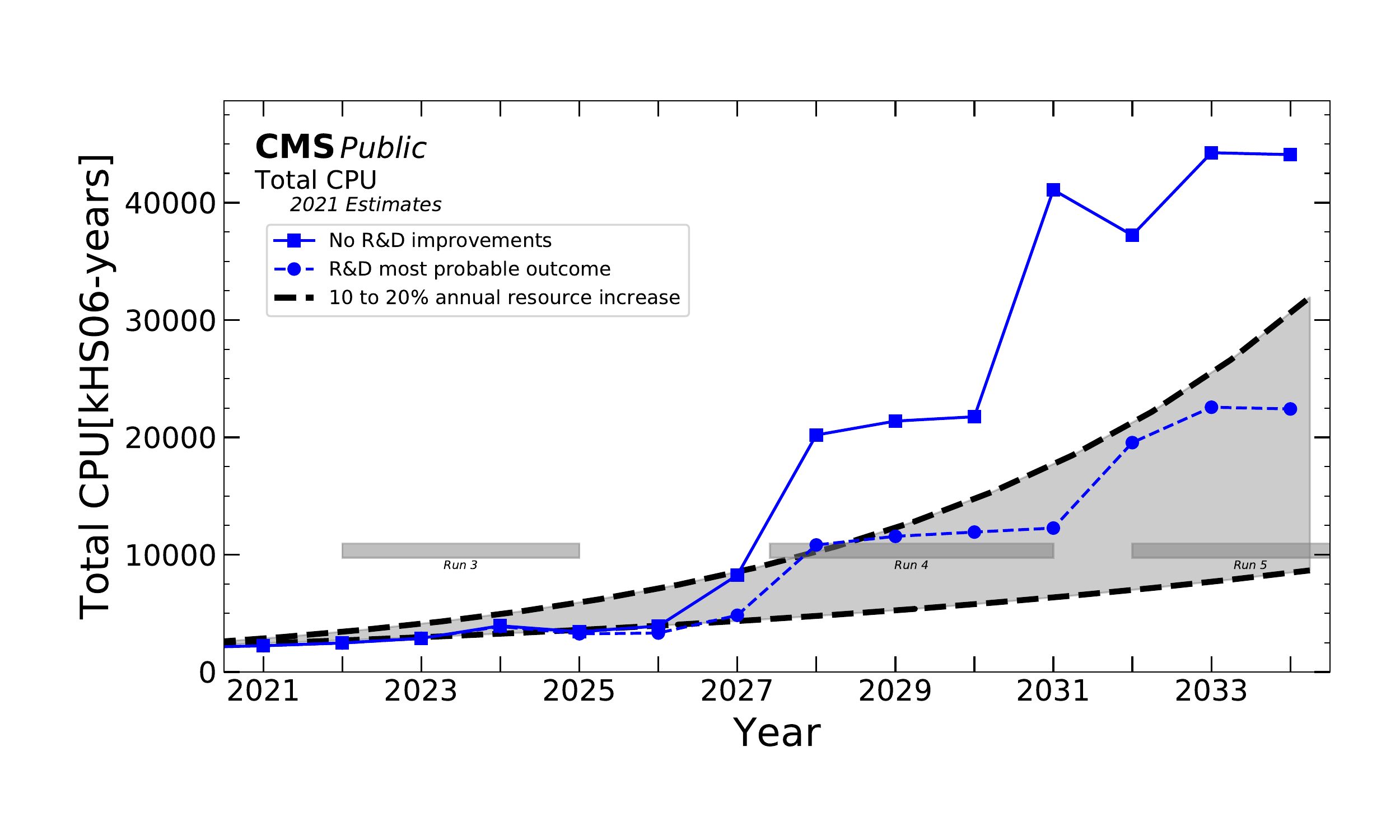}
    \end{subfigure}
    \begin{subfigure}[b]{0.8\linewidth}
        \centering
        \includegraphics[width=0.8\textwidth]{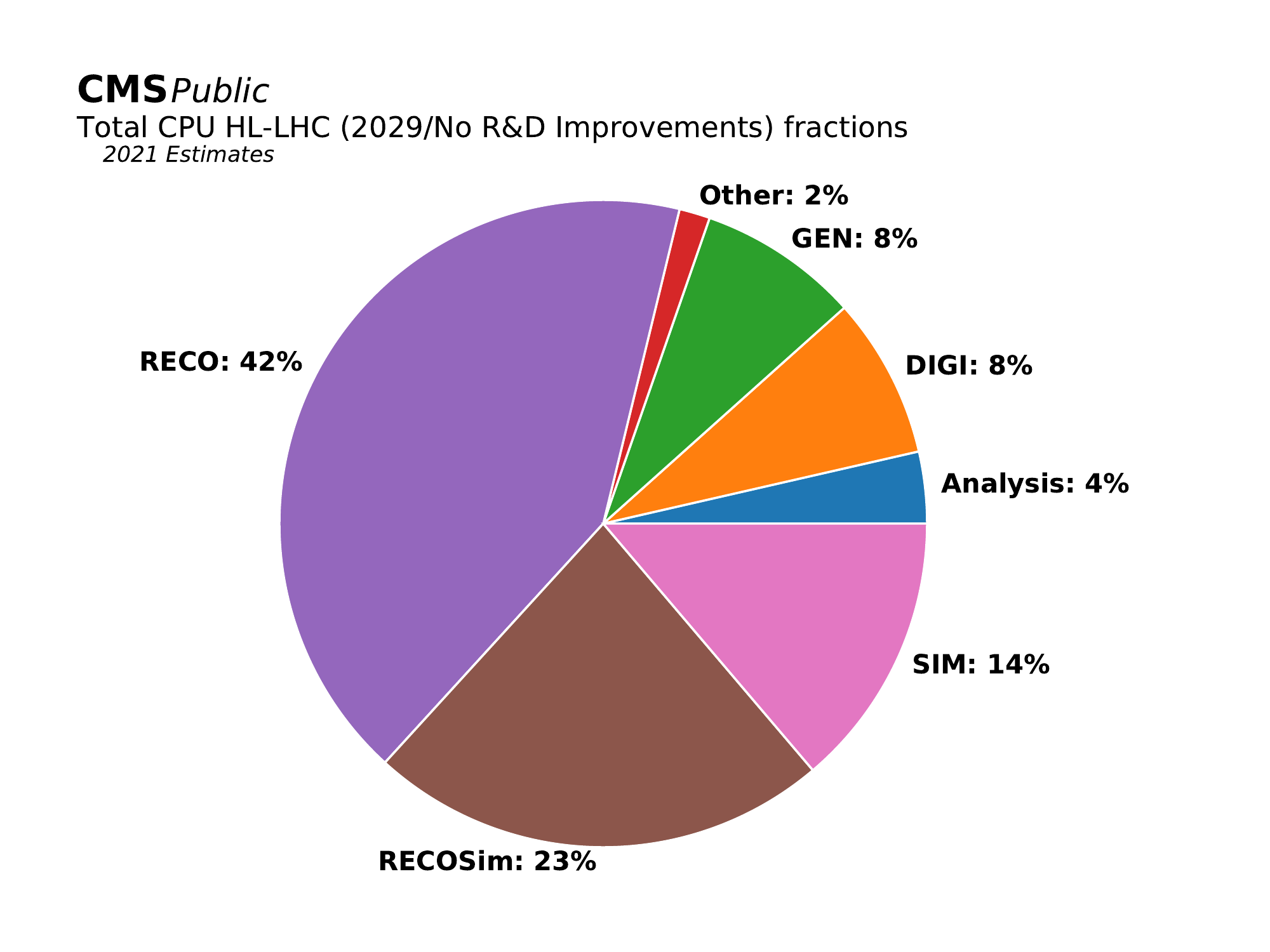}
    \end{subfigure}
    \caption{Approximate breakdown of CPU time, disk and tape requirements into primary processing and analysis activities for CMS during a typical HL-LHC year.}
    \label{fig:compute_resources}
\end{figure}

%-------------------------------------------------------------
A significant motivation for developing ML models for particle physics applications is the substantial computing and storage resource needs of experiments. Particularly, the planned high luminosity upgrade of the LHC (HL-LHC) will significantly increase both data density and processing complex; as shown in Figure \ref{fig:compute_resources}\footnote{source: \url{https://twiki.cern.ch/twiki/bin/view/CMSPublic/CMSOfflineComputingResults}}, the foreseen computing requirements cannot be met without significant R\&D efforts in algorithmic innovation and optimization, especially for data reconstruction and simulation pipelines. Graph-based representations and GNN architectures have shown substantial promise for a variety of particle physics tasks, including reconstruction and simulation, and in some cases have demonstrated better scaling properties, reduced resource utilization, more efficient data representations, and increased opportunity for parallelization and acceleration compared to traditional methods. However, several challenges remain to integrate these methods directly into experimental pipelines and reap the benefits of these innovations in practice. Furthermore, there difficulties specific to formulating particle physics problems in a graph-based framework and identifying appropriate architectures and evaluation methods. We outline the core challenges in applying GNNs in particle physics below.

\subsubsection*{Formulating HEP problems as graph problems}
It is not always straightforward to describe a particle physics problem in the formalism of graphs and GNNs. As described in Section \ref{sec:current_uses}, the same task can be formulated in multiple ways; for instance, the problem of charged particle tracking can be conceptualized as a graph edge classification task, an object condensation task, or an instance segmentation/object identification task. Each of these task types requires different graph construction, GNN architecture design, and training considerations. 
 
Graph construction determines the flow of information across the graph and allows for the representation and re-embedding of edge features. Due to the large size of detector datasets it is often not possible to form fully connected graphs and researchers must choose a graph construction method that respects the underlying physics of the data and the learning task. Several different graph construction methods have been explored for particle physics data including detector geometry based construction \cite{dezoort_charged_2021}, (dynamic) k-nearest-neighbors \cite{particlenet}, learned-representations \cite{qasim_learning_2019}, and set representations that avoid the need for edge construction altogether \cite{energyflow}. However, the impact of these graph construction methods on the downstream performance of a GNN model is not well characterized; we propose that further research on this subject and the exploration of additional graph construction algorithms will further enhance the expressivity of graph representations of particle physics data. 
 
Similarly, the specific GNN architecture used should be informed by the learning objective and the characteristics of the particle physics data. Although there is not a complete model of the impact of GNN architecture design on overall model performance, there are several key considerations that can be informed by knowledge of the underlying physics problem. For example, the decision of where to place the aggregation step of edge or node update blocks can be informed by the expected relationship between neighboring nodes; in \cite{DiBello:2020ppq} the authors assume that jet-tagging performance is affected by the $\Delta R$ between neighboring jets and so their architecture re-embeds graph edges then aggregates the representations for node updates while in \cite{energyflow} the authors re-embed each node representation independently and aggregate at the graph level. The choice of how many GNN modules to `stack' together in an architecture can be informed by the expected distribution of information across the graph; in addition to allowing higher-level representations of the data, multiple GNN modules also allow the exchange of information from progressively distant neighbors. Including attention mechanisms in a GNN architecture allows the model to learn to emphasize or deemphasize certain information during the aggregation step, for instance, allowing a jet tagging model to learn that secondary decay products are the most important signal for classification. The popular particle physics GNN architecture, GarNet \cite{qasim_learning_2019} uses a multi-headed self attention mechanism \cite{attention} to allow \textit{s} different node representations that are weighted and combined with a pre-defined, data-informed function. Finally, the choice of loss function is subject to several considerations such as whether or not the loss function preserves the symmetries of the data and the GNN architecture ( for example, the traditional MSE loss function is not invariant with respect to permutations of the output and targets). Additionally, in multitask architectures like \cite{thais2021instance} the balancing of task-specific loss terms becomes an important consideration.
    
In addition to implementation related concerns, there are important theoretical constraints and considerations. Unlike traditional feed-forward networks, message passing GNNs are not universal approximators. While precisely understanding the expressive power of GNNs is still an open question, recent work \cite{expressivity_functions, expressivity_limit,expressivity_survey} has begun to characterize their expressivity and potential for useful performance on different task types. In particular, on the task of graph isomorphism identification, \cite{expressivity_limit} demonstrated that the expressivity of message passing GNNs is at most equivalent to the Weisfeiler-Lehman heuristic. In the future, it is important that physicists are aware of these potential limitations to GNN expessivity and engage with graph theory and the broader ML research ecosystem to explore alternative GNN formalisms. 
    
\subsubsection*{Limitations of current tools} 

Compared to the well-established CV and NLP communities, tooling for graph-structured data is relatively immature. The most popular ML frameworks Pytorch~\cite{paszke2017automatic} and Tensorflow/Keras~\cite{tensorflow2015-whitepaper} have many built-in tools for CV and NLP tasks (e.g. TorchText and TorchVision) but do not natively support GNN approaches. Community extensions are making good progress, but there can be a steep learning curve, since the underlying frameworks are not tightly coupled with the extension (e.g. development and documentation may not be aligned). Aside from the direct task of GNN training and inference, there are many tangential tasks we may wish to perform on graph data. Those tasks which have historical significance in graph theory are generally supported, for example traversal and community finding \cite{hagberg2008exploring}. But these are not yet well-integrated into DL libraries, and are not always available on accelerators. Further, libraries that are optimized for sparse data on GPU may not be suitable for scientific data. For example, point cloud tasks often overlap with GNN-pipeline tasks, but libraries are often limited (or optimized towards) 3 dimensions \cite{DBLP:journals/corr/abs-2007-08501}. A scientific pipeline may need 4 dimensions (e.g. for spacetime vectors)~\cite{Wang_2021_WACV}, or even N dimensions (e.g. for operations in some learned latent space)~\cite{Ju:2021ayy}. 

Again, compared to CNN, RNN and dense graph (i.e. transformer) structures, ML optimization libraries for GNNs are immature. Many libraries that apply network pruning or quantization assume a model can be compiled to Onnx \cite{bai2019}. However, Onnx has not previously supported many of the operations required for message passing, whether the ``scatter reduce" operation in Pytorch, or ``unsorted sum" operation in TensorFlow. This has very recently been implemented, but these operations will need to be tested and carefully optimized for. Interpretability and explainability are notoriously slippery concepts, and offering libraries to capture them is already difficult in the scientific space, compared with mainstream ML. For example, there are many techniques for understanding the performance of CNN predictions on images, using Captum \cite{kokhlikyan2020captum}. However, this doesn't extend naively even to images in science - e.g. jet calorimetry image pixels are overlays of many particle charges. This is even more difficult in graphs, for which concepts of edge and shape feature filtering does not translate easily. Much work will need to be done in understanding even how to frame the question of interpreting and explaining GNN performance, let alone building general libraries that can apply to multiple scientific domains.
    
\subsubsection*{Speed of algorithms and memory footprint} 
The discrimination power of GNNs often stems from their ability to capture complex, sparsely-represented relational features that live on the edges - called ``anistropic message passing" above. This characteristic, which distinguishes GNNs from most other ML approaches, is also the most costly component of training and inference. In training, gradients must be back-propagated through edge features, meaning memory usage typically scales with number of edges, and with number of message passing steps~\cite{9355273}. While gradient checkpointing~\cite{DBLP:journals/corr/abs-1904-10631} can reduce the scaling with MP steps to O(1), the scaling with edges still remains a bottleneck for large graphs - as are typical in particle physics events. Some early solutions to this are discussed in the following section. Memory usage in inference is not such a hurdle where timing is not a priority. However, in the case of low-latency requirements, GNNs have been shown to work on FPGA systems with the HLS4ML library~\cite{DBLP:journals/corr/abs-2103-05579, Elabd:2021lgo}. Graph size on this hardware can be tuned between O(10) edges and O(1000), with latency scaling as the inverse to graph size. While this allows very low latency performance, in some applications graphs may be several orders larger, and indiscrimately sampling and distributing a graph across devices can affect a GNN's predictive capacity~\cite{NIPS2017_5dd9db5e}. Ultimately, many of the issues of memory and timing stem from the immaturity of sparse graph operations on accelerators. Many fused and memory-efficient operations that have been painstakingly designed for dense image data are still being redesigned for the sparse case. 
    
\subsubsection*{Integration of GNNs with full experiment pipelines} Many experiments are now making efforts to incorporate at least some concept of graph techniques or full GNN architectures in the data gathering and analysis pipeline~\cite{Qasim2021MultiparticleRI, Elabd:2021lgo, 2021APS..APRQ19007R}. Unlike some previous ML solutions for HEP phenomenology, graph-based ML is shown to benefit from incorporation of both low-level and high-level features~\cite{Wu2019ACS}, presumably due to its ability to hierarchically represent short and long-distance local information, as well as graph-level information. While this holds great promise for high-accuracy processing, it requires careful thought on how to integrate GNN-based pipelines across multiple stages of an experiment's dataflow. For example, the traditional ATLAS track reconstruction involves many hand-engineered stages that alternate between very low-level pixel charge information to reconstructed spacepoints to high-level calorimeter towers. Which of these being used depends on the hardware constraints and the goal of the particular stage. A GNN-based pipeline would benefit from the inclusion of \textit{multiple} levels of feature, which must be then propagated across heterogeneous computing devices.
Significant time and expertise are required to perform the same GNN operations across CPU, GPU and FPGA (as three example modalities for difference latency and budget requirements). And for the operations which are easily transferable, they are usually not optimized for the target hardware (e.g. graph construction on CPU vs. GPU). There are promising libraries that aim to provide modular operations for GNN and graph manipulation, and these will be discussed in the following section. A utopian situation would be a common HEP pipeline for graph-based analysis, but this is challenged by the very different data structures across experiment types. Even within an experiment, a sample may have heterogeneous types for the points of data (i.e. the nodes) and their relationships (i.e. the edges).

Within an experiment's GNN integration, there are sizeable differences in training and inference environments compared to industry and cloud environments. While historical motivation for GNNs in HEP came from computer vision (CV), where treating physics data as images has had great success, the implementation of GNNs in HEP fundamentally shares more similarity with the natural language processing (NLP) world. There, large language models (LLM) are built with transformers (conceptually equivalent to GNNs) and trained across many HPC nodes for many weeks. While training is prohibitively costly for many, models are then highly optimized for low memory/latency inference and implemented cheaply. The HEP community may be able to take inspiration from this behavior, where similar experiments could pool their resources to train large GNNs for complex physics tasks in the style of LLM (i.e. on a centralized cloud with as abstract data representations as possible). For each particular experiment, inference can then be optimized to the particular hardware, physics and budget constraints, and the general models transfer-trained (i.e. fine-tuned) towards an experiment's particular geometry. 

% [Example of tracking (graph building, edge classification, clustering, track fitting are all separate algorithms], [Needs heterogeneous computing],[Hardware/connection to detector design paper]

%\subsubsection*{Open source datasets} \TODO{related to benchmarking?, enabling collaaboration} include references to existing datasets (trackML, jet tagging, jetnet). not specifically for GNNs, what are limitations of these, what else do we need to see in this space. how do we benchmark to allow accurate comparisons. 
    
\subsubsection*{Collaboration across domains} In addition to the technical considerations outlined above, we also highlight that interdisciplinary collaboration is critical to the continued success of this research direction. In particular, in order for physicists to stay apprised of state-of-the-art innovations in graph representation methods and GNN architectures and theoretical developments in geometric machine learning, it is essential to build and maintain close collaborations with the broader ML and CS community. This community building is mutually beneficial as it can advance physics-informed and physics-inspired ML development and increase the broader research community's interest in and knowledge of particle physics computing problems and unique data considerations. While there are efforts in these areas (for example the ML and the Physical Sciences Workshop at NeurIPS\footnote{https://ml4physicalsciences.github.io}, the Physics Meets ML lecture series\footnote{http://www.physicsmeetsml.org/}, the Learning to Discover workshops and conference\footnote{https://indico.ijclab.in2p3.fr/event/5999/}, and others) there are more structural shifts required to maximize this information exchange and allow for more collaborative development. We point readers to Snowmass papers \textit{Broadening the scope of Education, Career and
Open Science in HEP} and \textit{Data Science and Machine Learning in Education} for more in depth discussion of cross-disciplinary community building considerations. 

GNNs have been widely utilized in many other domains of physical sciences as well. For example, GNNs have been successfully used for molecular \cite{molecule_property, molecule_property2} and chemical property prediction \cite{chemical_property}, protein structure and function prediction \cite{protein_function,protein_structure,alphafold}, and drug discovery \cite{drug_discovery,drug_discovery2,drug_discovery3}, among others. Many of these tasks are computationally similar to particle physics tasks and engaging with these research communities and others in the applied geometric machine learning community will likely yield widely beneficial advances.

%-------------------------------------------------------------
\section{Future Research Directions}
\label{sec:future}
%-------------------------------------------------------------
Considering the challenges outlined above, we propose that the community focus research and development efforts on methods incorporating existing physics knowledge and techniques into GNN architectures, expanding the space of tasks that GNNs are applied to, and developing new technique to increase the expressive power and ease of development of these models. We outline specific areas of focus, and their potential benefits, below. By defining a strong and complementary research trajectory we hope that the community will be able to 
maximize the potential benefits of applying graph-based data representations and GNNs to particle physics tasks.

\subsubsection*{Integrating physics knowledge into GNNs} 

As with all ML approaches, careful attention is being paid to how robust GNNs are to out-of-distribution performance. Especially in particle physics, where anomalies (whether genuine from new physics, or from some mis-calibration) are expected, any GNN-based pipeline must be shown to generalize well. A promising approach to this is including knowledge about the possible distribution in the GNN itself. This can be done in the structure of the data (e.g. including high level invariant features), the training of the data (e.g. augmenting with possible transformations or anomalies), or the architecture itself. This last is referred to as an ``equivariant", or symmetry-constrained, GNN. \cite{Bogatskiy:2020tje} showed that a GNN constrained to only perform convolutions that maintain Lorentz symmetry at each intermediate step is able to perform equally well when exposed to new, highly boosted jet data. There are early and very positive results in this burgening field, and we recommend the reader to review the Snowmass white paper \textit{Symmetry Group Equivariant Architectures for Physics}. Many of the equivariant architectures bearing fruit are graph-based, and as such that work is highly relevant to the ideas discussed here. 
    
%\TODO{equivariance, anything else?} Equivariant message passing networks (rotation, Lorentz, etc), CNNs do translation invariance, can computation be optimized for these kinds of invariances? refer to symmetry whitepaper. physics knowledge in graph construction (learned vs dynamic vs static) and model evaluation (specific metrics?) 

\subsubsection*{Data Augmentation} Increasing the diversity of data used in training machine learning models without actually collecting additional data has proven to be a powerful technique to improve model performance and generalization. Compared with other deep learning applications like image classification~\cite{DBLP:journals/corr/abs-1712-04621}, systematic studies of data augmentation for GNNs is rather limited. Data augmentation in GNN training has some unique challenges due to graph irregularity, although some of these can be mitigated with techniques such as utilizing neural edge predictors~\cite{DBLP:journals/corr/abs-2006-06830}.
    
In the context of GNN-based charged particle tracking for an LHC-like tracking detector, one approach which has shown promising results is to make a copy of each graph in the training set that has been reflected across the $\phi$-axis~\cite{ju_performance_2021}, where $\phi$ is the azimuthal angle of the track relative to the beam direction. The $\phi$ reflection creates the charge conjugate graph and helps to balance any asymmetry between positive and negatively charged particles within the training set. Similar studies for future GNN applications in HEP to improve model performance could be undertaken in the future.

\subsubsection*{Uncertainty quantification} 

The predictions of particle physics ML models are often not only used directly (e.g. for particle identification) but also feed into some eventual downstream calculation like a precision mass measurement of a SM particle or a distribution fit in a new particle search. Uncertainty in and ML model can directly limit the statistical significance of this final measurement. Thus, it is critical to consider not just the accuracy of a model's prediction, but also its uncertainty and robustness to overconfidence and out-of-distribution samples. 

Uncertainty quantification (UQ) is still an open area of research in ML, particularly for GNNs which have been shown to be sensitive to small perturbations in topology and node/edge features \cite{adversarial_gnn,topology_adversarial_gnn}. Generally, one must consider two types of uncertainty when characterizing the performance of an ML model: \textit{model} uncertainty describes how well a model's learned parameters fit the actual data distribution, while \textit{data} uncertainty describes uncertainty in the data distribution itself arising from noise in the data collection, data drift, adversarial attacks, statistical artifacts, or other sources. Additionally,the choice of loss function used to train a model implicitly defines a prior on the distribution of the residuals and can be considered another potential source of uncertainty or bias in a model that can limit the models ability to robustly estimate the underlying function and create high-fideity predictions for unseen data.

Much of the work on UQ for GNNs has focused on model uncertainty. Perhaps the simplest approach to UQ is model ensembling, where the same model is retrained multiple times with different random seeds and the variance across the model iterations is a proxy for the model uncertainty \cite{ensemble_uncertainty}. However, the bulk of recent work has centered on Bayesian methods that seek to characterize the posterior distribution of a model's decisions in order to obtain an estimate of the model's uncertainty \cite{bayesian_uncertainty}. A common approach to Bayesian UQ is the so-called `monte carlo dropout' method \cite{mc_dropout} that approximates the posterior distribution by collecting samples from dropout regularized forward passes of the model; however, there are also graph specific methods that perform Bayesian posterior updates independently for separate nodes or subgraphs \cite{graph_posterior}. 

Unfortunately, Bayesian UQ methods are often not scalable to real world scientific problems, and so several other UQ approaches have been developed in parallel, particularly by the scientific computing community. For example, orthogonal to Bayesian networks are Evidential Deep Learning (EDL) methods that formulate learning as an evidence acquisition process by modeling class probability predictions as a Dirichlet distribution which directly provides an estimate of the decision uncertainty \cite{EDL}; EDL has been applied with great success in molecular applications including property prediction and discovery \cite{edl_molecules}. The authors of \cite{surrogate_uq_materials} propose the use of surrogate models that exploit the benefits of GNNs for structured physical data while maintaining precise measures of model confidence; in this scheme a GNN is used to re-embed the data and form latent representations that are then used as inputs to an model like a Gaussian process that inherently provides UQ. The authors of \cite{uq_scienceml} demonstrate `learning by calibrating' by posing the estimation of prediction intervals as an additional learning task and training with a bi-level optimization formula.

Despite the importance of UQ for any particle physics application of ML, these methods have not been well explored for real world physics tasks. The most common method of UQ for particle physics is ensemble methods, but this has not been compared to other methods in terms of accuracy and efficiency in UQ computation, and in particular has not been well studied for GNNs specifically. In one of the few examples of these comparisons, the authors of \cite{uq_neutrinos} compare model ensembling (Naive Ensembling), monte carlo dropout, and EDL on three common neutrino physics reconstruction tasks including multi-particle classification using GNNs. They find that ensemble methods achieve the highest accuracy and the best calibration of output probability values. While this is promising, there needs to be substantial additional work comparing the performance of different UQ techniques across other physics datasets and tasks types. There may be additional benefits from improved UQ for GNNs; for example, \cite{uag} demonstrates that uncertainty estimates for individual nodes can be used to prevent information from noise or low quality data (like damaged or malfunctioning detector components) and \cite{surrogate_uq_materials} discusses potential uses of surrogate UQ models for experiment design. We encourage in particular the exploration of techniques beyond Bayesian UQ as those methods may be intractable for many physics applications. We also point readers to other Snowmass papers focused on UQ estimation for ML in particle physics, \textit{AI and Uncertainty Quantification} and \textit{Solving Simulation Systematics in and with AI/ML}.

\subsubsection*{Instance segmentation approaches} Standard message passing GNNs can only perform classifications at the node, edge, or graph level. Although these GNNs have been successfully applied to several key particle physics reconstruction tasks such as tracking, clustering, and jet building, they are typically used as an intermediate step rather than an end-to-end solutions. For example, in edge-classification-based tracking pipelines, the GNN weighted edges must be passed to a clustering algorithm to form track candidates that are then fit to extract final track parameters. In order to enable one-shot solutions to these reconstruction problems, it can be useful to re-conceptualize them as instance segmentation tasks. Instance segmentation is the task of detecting individual objects instances and their per-pixel segmentation mask, and is a critical task in computer vision. Recently, modified GNN architectures have been developed for 3D instance segmentation on point clouds. There have been two main approaches: predicting bounding shapes to separate individual object instances after re-embedding the graph through edge and/or node convolutions \cite{pointcloud_id,pointgnn} and (possibly attention based) graph pooling/clustering/condensation networks followed by localized node classification \cite{clustering_seg1,pointnet,clustering_seg2}. 
    
This type of GNN architecture has not been well studied for particle physics applications. As described in \cite{thais2021instance} there are challenges in adapting the boundary shape prediction models to particle physics data due to the irregular shape of physics objects (particularly tracks), however these approaches are still promising for enabling end-to-end reconstruction pipelines. Additionally, initial work has shown that object condensation approaches can successfully reconstruct particle clusters in calorimeters \cite{qasim_multi-particle_2021} and identify their originating particle type \cite{kieseler_2020_object}. We encourage the particle physics research community to consider what other tasks can be described with this framework and to continue exploring both GNN and non-GNN based instance segmentation architectures. In particular, reducing the number of separate algorithms needed for end-to-end GNN-based reconstruction could better allow these models to be used in computing resource constrained environments like the trigger system. 
    
\subsubsection*{Interpretability and explainability}

Many ML models are often referred to as `black boxes' as their large number of parameters and complexity of learned representations can prevent researchers from describing what information is relevant to a model's prediction. Interpretability and explainability are critical research areas in ML that seek to address this issue. There are several established explainability methods than can be easily modified to accommodate graph data and GNNs; these include sensitivity analyses on edge and node features, layer-wise relevance propagation within individual message passing networks, and disentangled representation learning that seeks to separate input information into latent features and encode them as separate dimensions that can then (ideally) be mapped back to human interpretable features. GNN attention mechanisms can also be considered as an interpretability method that allows model developers to identify relevant parts of a graph through the attention scores. 
    
Additionally, there are several explainability methods that have been developed specifically for GNNs. So-called `black-box approximation methods' train an inherrently interpretable model like linear regression or decision trees to approximate the GNN's decision boundary in the neighborhood of a specific target node and then use the interpretable model to explain the GNN decsion \cite{graphlime}. Similar to sensitivity analyses, perturbation-based explainability approaches seek to identify subgraphs that maximally contribute to the classification of an individual node or edge \cite{gnn_explainer,cf_gnn_explainer}. Graph filters adopt interpretability techniques originally applied to convolutional filters in CNNs to GNNs by incorporating graph kernels into the message passing process to identify relevant local graph structures \cite{kergnn}. 
    
Interpretability and explainability techniques are particularly relevant to physics applications. If an ML model outperforms a hand-tuned physics-driven model physicists are typically driven to characterize this difference in performance. Explainable GNN methods have not been widely adopted in particle physics, though there are a few examples. In \cite{explaining_mlpf}, the authors applied layerwise-relevance propagation to characterize the relevant nodes and features for a MLPF model while in \cite{symbolicreg_gnn} the authors apply symbolic regression as a form of disentangled representation learning to extract explicit physical relations including known force laws and Hamiltonians and a new analytic formula that can predict the concentration of dark matter from the mass distribution of nearby cosmic structures. There are also a few examples from the broader physical sciences space; in \cite{sa_materials} the authors use an Integrated Gradient method for sensitivity analysis to characterize the relationship between individual material grains and overall material property prediction and in \cite{counterfactual_molecules} the authors apply a counterfactual perturbation method to understand which components of molecules make them active for disease treatment.

We encourage physicists to invest in implementing and building upon these explainability methods for several reasons. The ability to precisely characterize the reasons for an ML model's downstream prediction greatly enhances the reliability, trustworthiness, and stability of the model and will better allow researchers to understand in which cases the model might be inaccurate so that they can create alternatives and fail-safes to prevent data loss and the introduction of additional uncertainty into measurements. By better understanding what information and operations are relevant to a model's predictions researchers may be able to build more efficient data representations, ML architectures, and even detectors; this is particularly relevant for GNNs where the data representation can correspond directly to the physical structure of the data. Additionally, particle physics in an extremely exciting area to explore interpretable and explainable ML techniques because so much of the physical theory underlying the data is already mathematically described. Thus, as shown in \cite{symbolicreg_gnn}, by mapping the behavior of ML models back into this mathematical language we can potentially uncover new physics and advance the state-of-the-art in explainable AI techniques.
    
\subsubsection*{New tools} 

There is a landslide of community-built libraries becoming available for graph-centric ML. As mentioned in the Challenges section, these are often modular and can take some effort to make compatible, compared with the CV and NLP tools that are included as first-class citizens in standard frameworks. Pytorch Geometric~\cite{Fey/Lenssen/2019}, DGL~\cite{DBLP:journals/corr/abs-1909-01315}, Graph Nets~\cite{DBLP:journals/corr/abs-1806-01261}, Jraph~\cite{jraph2020github} and Spektral~\cite{DBLP:journals/corr/abs-2006-12138} are a sample of the most popular kitchen-sink GNN libraries that run on Pytorch, JAX and/or TensorFlow. They vary in their focus, but generally they all attempt to cater to some of the highest hurdles to entry: a model zoo of implemented GNN operations, a framework and utilities for message passing and a method of batching graphs for training. 

There is still much work to be done in the surrounding graph manipulation technology, but some early implementations show that industry and scientific communities are moving to graph techniques in earnest. For example, RAPIDS AI~\cite{RAPIDSAI} dedicates a whole sublibrary to accelerated graph techniques, called cuGraph. N-dimensional point cloud libraries are also being made available for both CPU and GPU, for example for fast graph construction with exact nearest neighbors~\cite{xue2022}, and approximate nearest neighbors~\cite{johnson2019billion}. Optimization of GNN operations is also receiving much attention from industry. As mentioned in the Challenges section, Onnx supports GNN operations, and these will soon be aligned with the major frameworks. TensorRT~\cite{tensorrt}, a popular library for optimizing CNN-based networks, is collaborating with several scientific groups to extend these optimizations to GNNs, and this has already been done in the form of fused operations in the DGL library. Finally, the recently release Open Graph Benchmark (OGB) \cite{hu2020ogb,hu2021ogblsc} fills a significant gap that traditionally CV and NLP have prioritized: uniform and translatable model evaluation. This library is compatible with both PyG and DGL and should significantly speed up comparisons between GNNs, a task that should be compulsory for any new implementation on a physics use-case but that is typically laborious and inconsistent.
    
\subsubsection*{New task types} 

The majority GNN applications in HEP focus on well-studied reconstruction and identification tasks. Recent paradigms emerging in the HEP community, for example anomaly detection methods for model-agnostic physics searches \cite{kasieczka_lhc_2021}, present an exciting opportunity to develop novel graph-based algorithms. To date, several studies have demonstrated the applicability of autoencoders applied to particle graphs for anomaly detection \cite{atkinson_anomaly_2021, tsan_particle_2021}. %\TODO{add a bit more here...heterogenous graphs? manifold learning? full event characterization?}
    
\subsubsection*{New GNN operations} 

It is difficult to predict the direction that graph neural network research will take, but we can be informed both by the trajectory of CNNs and transformers (which serve as inspiration and motivation for much GNN development), as well as by the typical tasks in scientific analysis. A clear research trend is towards heterogeneous data types, whether node, edge or whole-graph - closely related to "multi-modal" learning~\cite{DBLP:journals/corr/BaltrusaitisAM17}. An example use-case may be differing volumes within a detector, so that rather than a single MLP within a GNN being expected to handle multiple sets of features trivially combined, different layers of the GNN may learn natively on the various input features. This could be extended to the case of a GNN applied simultaneously to sensors that produce numerical outputs and elsewhere sensors that output image-like data, akin to vision transformers. These operations are now natively supported in Pytorch Geometric, for example. 

Hierarchical features have proven to be very important in CV research, for example as extracted by pooling in CNNs. Edge contraction, node pooling, and hypergraph learning may be one direction to emulate this high performance \cite{DBLP:journals/corr/abs-2111-00180, DBLP:journals/corr/abs-1910-11436, DBLP:journals/corr/abs-1806-08804}. However, many GNN operations are limited by the WL isomorphism test mentioned in the Challenges section. Research is underway for generalisations to these WL-limited GNN operations. Hierarchical features may be able to be recovered by including information about neighborhood structure~\cite{DBLP:journals/corr/abs-2006-09252}, or pooling according to high-importance connected edges. 

These ideas will ultimately feed into more powerful graph generation techniques. Hierarchical features have long been a staple in image generation, for example in StyleGAN~\cite{DBLP:journals/corr/abs-1812-04948} and Progressive GANs~\cite{DBLP:journals/corr/abs-1710-10196}, where high fidelity images are generated by relying on different layers to account for different levels of granularity. Without having GNN operations that can capture hierarchical information (and that are not prohibitively expensive to compute), graph generation may be confined to the same capabilities as non-hierarchical CNNs, which struggle to generate even MNIST~\cite{cheng2020analysis}.
    
%\TODO{Could also include ways of improving GNN expresivity like subgraph operations https://towardsdatascience.com/using-subgraphs-for-more-expressive-gnns-8d06418d5ab}

%-------------------------------------------------------------
\section{Conclusions}
\label{sec:conclusions}
%-------------------------------------------------------------
ML has become an integral part of particle physics and continued research and development in these areas will likely be necessary to maximize the physics potential of current experiments and to meet the computing requirements of future experiments. Amongst the data structures currently utilized in ML, graphs are the most intuitive representation for variable size, geometrically structured particle physics data. The adoption and development of GNNs for physics tasks has proven extremely successful over the past several years and given the expressive power of these models it is likely that this success will continue to grow. 

Given their ability to handle sparse data and the irregular geometries of some particle detectors, we propose that GNNs should be a universal benchmark for key particle physics tasks. In other words, when new ML or traditional physics motivated approaches are developed for a reconstruction or simulation task, they should be directly compared to a graph (or set/pointcloud) based approach. However, this requires continued  or expanded support and investment from the research community; we particularly encourage work on the development of relevant software tools for GNN model building and that experiments ensure that their software pipelines are amenable to the incorporation of these models.

%\TODO{What do we want to say here} GNNs should be the default for HEP given their expressivity. Give the irregularity of detector geometry/data sparsity etc GNNs should always be a benchmark compared to 'hardcoded' models. More support and investment in development (particularly in experiment software) 

\def\thefootnote{\fnsymbol{footnote}}
\setcounter{footnote}{0}

% Bibliography

\bibliographystyle{JHEP}
\bibliography{references}

\end{document}